\documentclass[10pt]{article}
\usepackage{epsfig}
\usepackage{feynmf}

\begin{document}

\title{Pion Decay Widths of $D$ mesons}

\author{K.O.E. Henriksson,$\,$\footnote{koehenri@pcu.helsinki.fi}$\:$
T.A. L\"{a}hde,$\,$\footnote{talahde@pcu.helsinki.fi}$\:$
C.J. Nyf\"{a}lt$\,$\footnote{nyfalt@pcu.helsinki.fi}$\:$ and
D.O. Riska$\,$\footnote{riska@pcu.helsinki.fi}}

\date{}
\maketitle

\centerline{\it Department of Physics}
\centerline{\it and}
\centerline{\it Helsinki Institute of Physics, POB 9, 00014 University of
Helsinki, Finland}

\thispagestyle{empty}
\vspace{1cm}

\begin{abstract}
The pionic decay rates of the excited $L=0,1$ $D$ mesons
are calculated with a Hamiltonian model within the framework of the
covariant Blankenbecler-Sugar \mbox{equation.} The interaction between
the light quark and charm antiquark is described by a linear scalar
confining and a screened one-gluon exchange interaction. The decay widths
of the $D^*$ mesons obtain a contribution from the exchange current that
is associated with the linear scalar confining interaction. If this
contribution is taken into account along with the single quark
approximation, the calculated decay rates of the charged $D^*$ mesons are
readily below the current empirical upper limits if the axial coupling
constant of the light constituent quarks is taken to be $g_A^q$ = 0.87,
but reach the empirical upper limits if $g_A^q$ = 1. With the conventional
values for $g_A^q$, the calculated widths of the $D_1$ and $D_2^*$ mesons
fall somewhat below the experimental lower limits, leaving room for other
decay modes as well, such as $\pi\pi$ decay. The unrealistically large
contribution from the axial charge operator to the calculated pion decay
width of the $D_1$ meson is suppressed by taking into account the exchange
charge effects that are associated with the scalar linear confining and 
vector one-gluon exchange interactions. The predicted values for
the pionic widths of the hitherto undiscovered $L=1$ $D_1^*$ and $D_0^*$
mesons are found to be smaller than previous estimates.

\end{abstract}
\newpage

\section{Introduction}

The pion decay rates of the excited charmed mesons - the $D$ mesons - may 
provide direct information on the strength of the pion coupling to
light constituent quarks. As the charm quark in the $D$ mesons does not
couple to pions, the decay mechanism is determined by the pion coupling
to the light flavor constituent quark. A first assumption is that this
coupling is independent of the interaction between the light quark (or
antiquark) and the charmed antiquark (or quark). While this may be
considered as a satisfactory approximation for the axial current part of
the pion-quark coupling, it leads to large overestimates of the decay
widths in the case of the axial charge term, a problem that may be cured
by the two-quark mechanism that is associated with the confining
interaction between the light and charmed quark in the $q\bar c$ $(\bar q
c)$ system.

Because of the large velocities of the confined quarks in $D$ mesons,
and in view of the small mass of the light constituent quarks, the
$D$ mesons have to be described as relativistic interacting two-particle
systems. The simplest way to achieve such a description is to employ a
covariant three dimensional reduction of the
Bethe-Salpeter equation and a corresponding quasipotential
representation of the interaction between the constituents.
The hitherto considered quasipotential
descriptions for the $D$ mesons are based on the Gross~\cite{Gross} and
the Blankenbecler-Sugar equations~\cite{Blank,Logu}. Both approaches
have been shown to yield reasonable predictions for the $D$ meson
spectrum with combinations of linear confining and one-gluon exchange
interactions between the quark and antiquark~\cite{Orden,Ebert,Lahde}.

This dynamical model, with static interactions 
in the $q\bar Q$ system applied in
the solution of the Gross equation, has recently been used to calculate
the pseudoscalar meson decay rates of the $D$ mesons~\cite{Goity}. A
fair description of the decay rates of the $D^*$ mesons was achieved,
under the assumption that the pions couple to the light constituent quarks
by the standard (pseudovector) coupling. 
On the other hand, the ratio of the widths of the $D_2^*(2460)$ and
the $D_1(2420)$ mesons was found to be $\sim $ 2.5, while the 
empirical ratio is $\sim 1.3$.
In view of the importance
of determining the form and strength of the pion-quark coupling, an
analogous calculation within the framework of the Blankenbecler-Sugar
equation is performed here. There is no obvious reason for preferring
one or the other quasipotential framework, besides that of calculational
convenience. The Gross equation framework has the virtue of
reducing to a Dirac equation for the light quark in the infinite mass
limit of the heavy (charm) quark. The Blankenbecler-Sugar equation has
an interpretational advantage in its formal similarity to the standard
Schr\"odinger equation framework.

To calculate the pion decay rates of the excited $D$ mesons, the 
wavefunctions that have been obtained in ref.~\cite{Lahde} by solving the
Blankenbecler-Sugar equation in configuration space are employed in order
to achieve a description of the states in the $D$ meson spectrum. These 
wavefunctions correspond to a model which
describes the interaction in the $q\bar Q$ system as a scalar linear
confining interaction combined with a screened relativistic one-gluon
exchange interaction, which yields a spectrum that agrees with the
empirically known part of the spectrum. The model leads to hyperfine
splittings that agree well with recent NRQCD lattice calculations in the
quenched approximation~\cite{Wolo}. The rates for the decays
of the form $D'\rightarrow D\pi$ are then
obtained by calculating the matrix elements of the pion creation
operator between the excited and ground states using such wave
functions. Because of the small mass of the light flavor quark, which
couples to the pions, the non-local structure of the pseudoscalar
coupling of the pion to the quark has to be treated in unapproximated
form. The static (local) approximation to this vertex function is shown
to imply an overestimate by about a factor 2.

The present empirical information on the widths of the excited charm
mesons remains very incomplete. Absolute values, with large uncertainty
ranges, are known for the decay widths of the $D_1(2420)$ and
$D_2^*(2460)$ mesons, but for the $D^*(2010)^\pm$ and the
$D^*(2007)^0$ only upper limits are available at present~\cite{PDG}. The
pion decays of the $D^*$ mesons to $D$ mesons are $P-$wave
decays generated by the axial current operator.
If the pion is assumed to be emitted by a single quark
operator with the conventional value
for the pion-quark coupling constant, the calculated
widths for the decays $D^*\rightarrow D\pi$ fall well
below the present empirical upper limits in the case of the
charged $D^*$ mesons (The empirical upper limit on the
width of the neutral $D^*$ meson is too large to be constraining).
Upon addition of the contribution from the two-body axial exchange
(pair) current that is associated with the linear scalar 
confining interaction, the calculated widths reach the
empirical upper limits in case of the charged $D^*$ mesons,
if the value of the axial coupling constant of the light
constituent quarks is taken to be $g_A^q=1$.

The single quark mechanisms for pion production 
lead to a considerable overprediction of the $S$-wave pion decay
widths of the $D_1$ mesons. This overestimate may be reduced by invoking 
the two-quark mechanism, which is naturally associated with the scalar
confining and vector one-gluon exchange interactions. Consequently, the
predicted widths of other $D$ meson states that decay by an $S$-wave
mechanism are also suppressed by large factors. This effect was first
hinted at in
ref.~\cite{Goity}. In the case of the scalar confining interaction the
simplest description of this mechanism is to view it as an effective
increase of the constituent quark mass from $m_q$
to $m_q+cr$, where $c$ is the confining string tension. Since the quark
mass appears in the denominator of the transition amplitude for $S$-wave
pion decay, this increase of the constituent quark mass leads to a large
reduction of the associated matrix element. An analogous suppression of 
$S-$wave pion decay modes was achieved in ref.~\cite{Goity}
as a correction to the one-quark operator through coupling
to negative energy states. This two-body mechanism is analogous to
that, which is required for a realistic description of the M1 decays of
charmonium and heavy light mesons~\cite{Lahde,Lahde1}.

Once this two-quark mechanism is taken into account, the
calculated pion decay widths of the $D_1(2420)$ and
$D_2^*(2460)$ mesons fall somewhat below
the empirical values if the axial coupling of the constituent
quark is taken to be less than 1. If, on the other hand, the 
matrix element of the pion decay amplitude is evaluated in
the non-relativistic approximation, the calculated decay
widths exceed the empirical values, in agreement with the
result of ref.~\cite{Goity}. In the case of the $D_1$ meson, the $S-$wave
pion decay mode is found to contribute significantly, so that in the end
the net ratio of the calculated widths of the $D_2^*$ and $D_1$ mesons
is about 1.2, which falls within the wide uncertainty range
of the current experimental value 1.3.
The underprediction of the pionic decay widths of the
$L=1$ charm mesons is natural, as a substantial fraction of the
total width is expected to be due to other decay modes, in
particular $\pi\pi$ decay. The analogy with the corresponding decay
modes of the
$K_2^*(1430)$ strange meson suggests that the decay modes $D'\rightarrow
D\pi\pi$ may be responsible for a significant fraction of the observed
decay widths.

The results for the calculated pion decay widths of the excited $D$ mesons
obtained here are rather similar to those
obtained in ref.~\cite{Goity}, despite the different calculational
framework and the different Hamiltonian model. 
The calculation in ref.~\cite{Goity} was restricted to the decays allowed
by the lowest order selection rules suggested by heavy quark
symmetry~\cite{Isgur}. In the case of the $D_1$ meson the excluded
$S-$wave transition was found to be rather important here, a result
already hinted at in ref.~\cite{Mehen}. The conclusion reached here is
that the chiral quark model does indeed provide a fair description of the
pion decay widths of the orbitally excited $D$ mesons, if they are treated
as relativistic interacting two-quark systems.

This paper falls into 5 sections. In section 2 the decay width
for $D^*\rightarrow D\pi$ is calculated. In section 3 the
corresponding decay widths for the orbitally excited
$D_1$ and the $D_2^*$ mesons, including the $S$-wave pion decay rate of
the $D_1$ meson, are calculated.
In section 4 the estimated pion decay widths of the hitherto
undiscovered charm mesons with $L=1,J=1$ and $L=1,J=0$
are given. 
Section 5 contains a summarizing discussion.


\section{Pion decay widths of the $D^*$ mesons}

\subsection{Single quark approximation}

The main contribution to the decay widths of the $D^*$ mesons in the
ground state band is due to the pion decays $D^*\rightarrow
D\pi$. These transitions are intriguing in that since the mass
difference $M_{D^*}-M_D$ is very close to the pion mass, the available
phase space is very small. Because of this closeness to the threshold
for $\pi$ decay, and the nonzero mass splittings between the different
charge states of the $\pi$ and $D$ mesons, one of these decays -
the decay of the $D^{*0}$ to $D^\pm \pi^\mp$ - is in fact kinematically
forbidden. The orbital wave functions of the constituents of the $D$ and
$D^*$ mesons differ very little from one another, which implies that the
main pionic decay mechanism is $P$-wave pion decay. Because of the
consequent threshold suppression and the small phase space, the total
widths of the $D^*$ mesons are expected to be very small. The current
empirical upper bound for the total width of the $D^{*\pm}$ is 0.131 MeV
and that for the $D^{*0}$ is \mbox{2.1 MeV~\cite{PDG}.} The former one of
these upper bounds is already constraining for theoretical model
calculations.

\begin{figure}[h!]
\begin{center}
\epsfig{file = 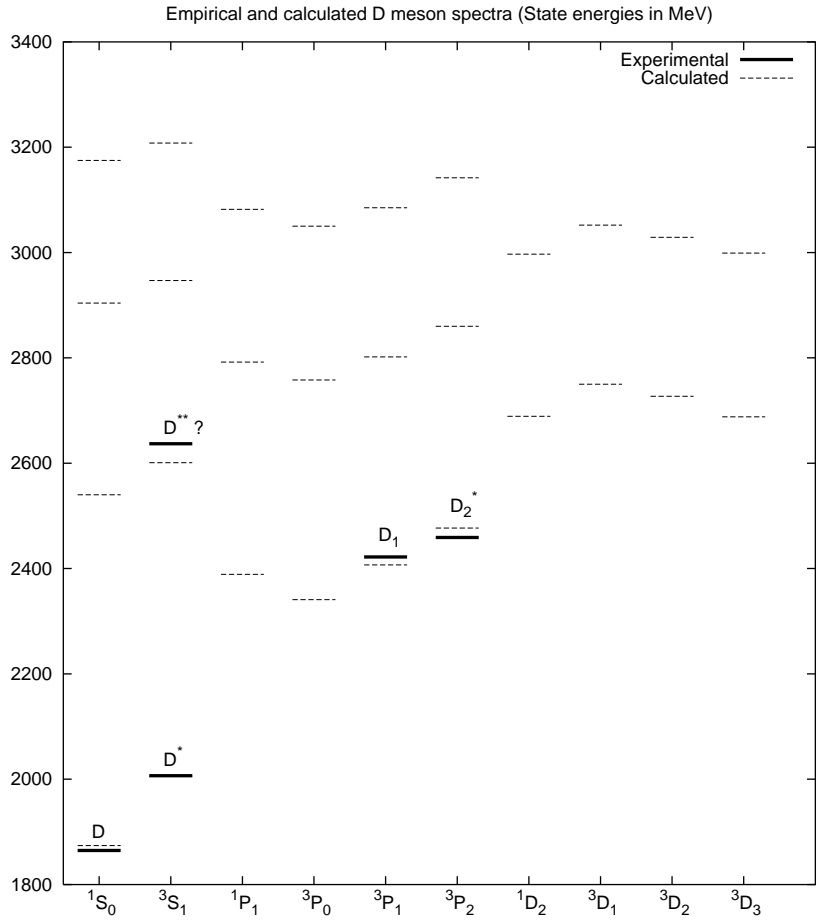}
\caption{Experimental and calculated $D$ meson states. The calculated
states are solutions to the Blankenbecler-Sugar equation and correspond to
the Hamiltonian used in ref.~\cite{Lahde}. The energies of the empirical
states along with the pion momenta used in the calculations are listed in
Table~\ref{masstab}.}
\label{spektr}
\end{center}
\end{figure}

\begin{figure}[h!]
\begin{center}
\epsfig{file = 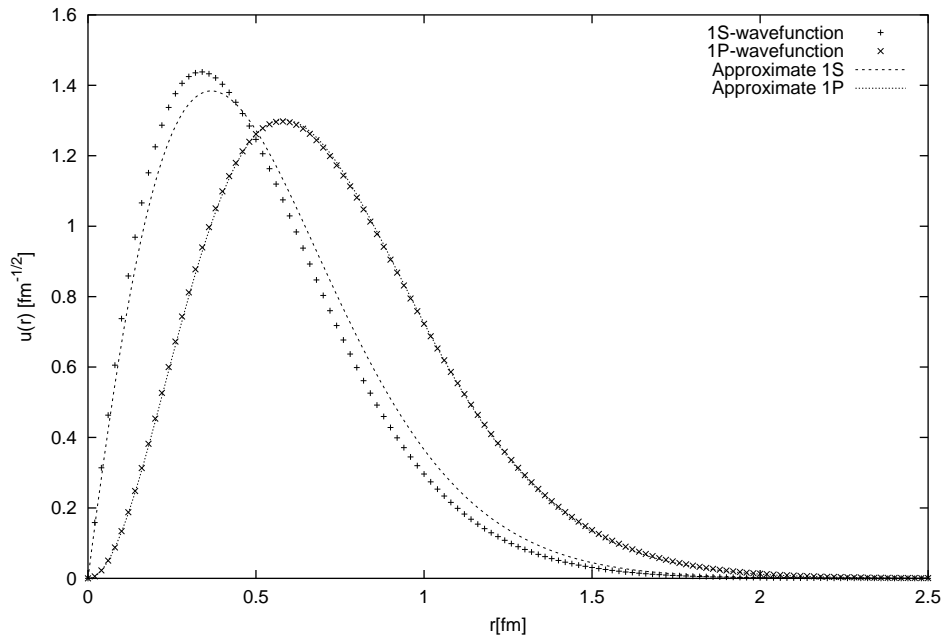}
\caption{The reduced wave functions for the $S$- and $P$-state $q\bar c$
(or $\bar q c$) systems according to ref.~\cite{Lahde}. As these
wavefunctions correspond to the spin-averaged $S$- and $P$-states of the
$D$ meson, they do not incorporate the fine structure splittings of the
various states in Fig.~\ref{spektr}. Therefore, in order to avoid
unnecessary errors, the masses and pion momenta listed in
Table~\ref{masstab} are used in the calculations. The approximate
wavefunctions corresponding to eq.~(\ref{appr}) in Appendix~A are shown
for comparison.}
\label{wavefig}
\end{center} 
\end{figure}

The $D$ and $D^*$ mesons are confined $q\bar c$ and $\bar q c$ systems,
where $q$ and $\bar q$ denote constituent $u,d$ and $\bar u,\bar d$
quarks and $c$ and $\bar c$ charm and anticharm quarks
respectively. The pions only couple to the light flavor quarks,
the simplest model for the coupling being the chiral coupling:

\begin{equation}
{\cal L}=i{g_A^q\over 2f_\pi}\bar\psi_q\,\gamma_5\gamma_\mu
\,\partial_\mu\,\vec\phi_\pi\cdot\vec \tau\,\psi_q.
\label{lagr}
\end{equation}
Here $g_A^q$ denotes the axial coupling constant of the light flavor
constituent quarks and $f_\pi$ is the pion decay constant (93 MeV).
The value of $g_A^q$ should be somewhere between
unity~\cite{Weinberg1} and 
$g_A^q=0.87$~\cite{Weinberg,Dicus,Dannbom}. Only the $P$-wave part of the
coupling~(\ref{lagr}) contributes to the pionic decays $D^*\rightarrow
D\pi$, and hence, in the single quark approximation, the transition
operator reduces to

\begin{eqnarray}
T_P&=&{g_A^q\over 2f_\pi}\:\bar u(\vec p\,')\:
\gamma_5^q\vec \gamma^q\cdot \vec k\:u(\vec p\,)\:\tau_\pi
\nonumber \\
&=&-i{g^q_A\over 2f_\pi}\sqrt{{E'+m\over 2E'}}\sqrt{{E+m\over
2E}}\left(1-{P^2-k^2/4\over 3(E'+m)(E+m)}\right)\vec\sigma^q\cdot\vec
k\,\tau_\pi.
\label{oper}
\end{eqnarray}
Here $m$ denotes the mass of the light constituent quark and
$\vec k$ is the momentum of the emitted pion. The operator
$\vec P$ is defined as
$(\vec p\,'+\vec p\,)/2$, with $\vec p$ and $\vec p\,'$ being the initial
and final momenta of the light quark. The energy factors $E$ and $E'$
that appear in eq.~(\ref{oper}) are defined as $\sqrt{p^2+m^2}$ and
$\sqrt{p'^2+m^2}$ respectively. The pionic decay widths of the $D^*$
mesons may now be expressed as

\begin{equation}
\Gamma\left(D^{*0}\rightarrow D^0\pi^0\right) = {1\over
24\pi}{E_{D^0}\over M_{D^{*0}}}\left({g_A^q\over f_\pi}\right)^2k^3 
{\cal M}_0^2
\end{equation}
for the $D^{*0}$ meson, and

\begin{eqnarray}
\Gamma\left(D^{*\pm}\rightarrow D^{\pm}\pi^0\right)&=&{1\over
24\pi}{E_{D^{\pm}}\over M_{D^{*\pm}}}\left({g_A^q\over f_\pi}\right)^2 k^3
{\cal M}_0^2, \\
\Gamma\left(D^{*\pm}\rightarrow D^0\pi^{\pm}\right)&=&{1\over
12\pi}{E_{D^0}\over M_{D^{*\pm}}}\left({g_A^q\over f_\pi}\right)^2 k^3
{\cal M}_0^2
\end{eqnarray}
for the charged $D^*$ mesons. In the above expressions, ${\cal M}_0$ is
the orbital part of the matrix element
$\left<00,00\right| \vec\sigma_q\cdot\vec k \left|01,1m\right>$, where
$\left|LS,JM\right>$ denotes the state vector of the $q\bar Q$ system. To
arrive at this expression, the following result for the sum
over spins and integration over directions of $\vec k$ has been used:

\begin{equation}
\frac{1}{3}\int d\Omega_k \sum_{m}\left<1m\right|\vec
\sigma_q\cdot\vec k\left|00\right>\left<00\right|\vec\sigma_q\cdot
\vec k\left|1m\right>={4\pi\over 3}k^2.
\end{equation}
Here $\vec \sigma_q$ denotes the spin operator of the light constituent
quark. The matrix element ${\cal M}_0$ may be expressed in
unapproximated form as

\begin{eqnarray}
{\cal M}_0&=&{1\over
\pi}\int_{0}^{\infty}dr'r'u_0(r')\int_{0}^{\infty}dr\,r\,
u_0(r)\int_{0}^{\infty}dP\,P^2\int_{-1}^{1}dz\,f_{\mathrm{BS}}(P,z) 
\nonumber \\
&&\sqrt{E'+m\over 2 E'}\sqrt{E+m\over 2 E}
\left(1-{P^2-k^2/4\over 3(E'+m)(E+m)}\right) \nonumber \\
&&j_0\left(r'\sqrt{P^2+{k^2\over
16}+{Pkz\over 2}}\:\right)j_0\left(r\sqrt{P^2+{k^2\over 16}-{Pkz\over
2}}\:\right).
\label{matrix}
\end{eqnarray}
Here $u_0(r)$ is the reduced radial wave function for the $D$ and $D^*$
mesons. The explicit expressions for the energy factors appearing in
eq.~(\ref{matrix}) are 

\begin{equation}
E=\sqrt{m^2+P^2-Pkz+k^2/4}\quad\mathrm{and}\quad
E'=\sqrt{m^2+P^2+Pkz+k^2/4}
\end{equation}
respectively. The factor $f_{\mathrm{BS}}(P,z)$ originates in the
quasipotential reduction of the amplitudes defined for the Bethe-Salpeter
equation, and is defined as

\begin{equation}
f_{\mathrm{BS}}(P,z)= \frac{M+m}{\sqrt{(E+E_c)(E'+E_c')}}.
\label{bfact}
\end{equation}
Here $E_c=\sqrt{M^2+P^2-Pkz+k^2/4}$ and $E'_c=\sqrt{M^2+P^2+Pkz+k^2/4}$
and $M$ is the heavy quark mass. In the non-relativistic limit the
expression~(\ref{matrix}) reduces to

\begin{equation}
{\cal M}_0=\int_{0}^{\infty} dr\,u_0^2(r)\,j_0\left({kr\over 2}\right).
\label{nonrel}
\end{equation}
Because of the small masses of the light constituent quarks
the non-relativistic approximation
~(\ref{nonrel}) turns out to be inadequate for a reliable description,
and typically leads to overestimates of the exact result obtained from
eq.~(\ref{matrix}) by at least a factor $\sim 2$. In order to evaluate the
expression~(\ref{matrix}), a model for the wavefunctions is required. In
this work, the wavefunctions that were obtained by solving the
Blankenbecler-Sugar equation in configuration space in
ref.~\cite{Lahde} will be used. Those wavefunctions correspond to an
interaction Hamiltonian which is a combination of relativistic one-gluon 
exchange and a scalar confining interaction. In this model the light quark
mass was obtained as 450 MeV and that of the charm quark as 1580 MeV. The
string tension in the scalar confining interaction was taken to be
\mbox{$c$ = 1120 MeV/fm}. The relativistic one-gluon exchange interaction
employed in ref.~\cite{Lahde} also features a running color "couplant",
which was taken to be of the form~\cite{Matting}:

\begin{equation}
\alpha_s(k^2)={12\pi\over 27}{1\over \ln\left[{k^2+4m_g^2\over
\Lambda_0^2}\right]}.
\label{zzz}
\end{equation}
In eq.~(\ref{zzz}), $\Lambda_0$ denotes the confinement scale, for
which the value 280 MeV was used in ref.~\cite{Lahde}. For the gluon mass
parameter $m_g$, the value $m_g$ = 240 MeV was obtained. A comparison
between the spectrum obtained in ref.~\cite{Lahde} and the current
empirical spectrum is shown in Fig.~\ref{spektr}. The resulting reduced
radial wave functions for the $1S$- and $1P$-states that were obtained
with this model in ref.~\cite{Lahde} are shown in Fig.~\ref{wavefig}. 

\newpage

\begin{table}[h!]
\begin{center}
\begin{tabular}{l|c|c|c}
\quad$\:\:$ Decay & Initial state mass ($M_i$) & Final state mass
($M_f$) & $\pi$ momentum ($k$) \\ \hline\hline &&& \\ 
$D^{*0} \rightarrow D^0{\pi}^0$       & 2007 & 1865 & 43.1 \\
$D^{*\pm} \rightarrow D^{\pm}{\pi}^0$ & 2010 & 1869 & 38.3 \\
$D^{*\pm} \rightarrow D^0{\pi}^{\pm}$ & 2010 & 1865 & 39.6 \\
&&& \\ 
$D_1 \rightarrow D^*\pi$	      & 2422 & 2009 & 355  \\
$D_2^* \rightarrow D\pi$	      & 2459 & 1867 & 505  \\
$D_2^* \rightarrow D^*\pi$	      & 2459 & 2009 & 389  \\
$D_1^* \rightarrow D^*\pi$	      & 2389 & 2009 & 326  \\
$D_0^* \rightarrow D\pi$	      & 2341 & 1867 & 408  \\
\end{tabular}
\caption{Initial and final state $D$ meson masses and emitted pion momenta
in MeV used in the calculations. When different charge states are
specified, the values are taken from~\cite{PDG}, otherwise suitable
averages are used instead. The $D_1^*$ and $D_0^*$ masses are taken from
ref.~\cite{Lahde}.}
\label{masstab}
\end{center}
\end{table}

The numerical value of the unapproximated matrix element ${\cal M}_0$,
eq.~(\ref{matrix}), as obtained using the wave function $u_0(r)$ shown in
Fig.~\ref{wavefig} is ${\cal M}_0 = 0.649$. Although the value of the pion
momentum $k$ in the decays of the $D^*$ mesons is
non-negligible (In the case $D^{*\pm}\rightarrow D^0\pi^+$ it is
\mbox{39.6 MeV/c} and in $D^{*0}\rightarrow D^0\pi^0$ \mbox{43.1 MeV/c})
the product of this momentum and the range of the wave function $(\sim
0.5\:\mathrm{fm})$, is of the
order $\sim 0.1$. Because of the smallness of this product the value of
the non-relativistic approximation to the matrix element
${\cal M}_0$ as evaluated from eq.~(\ref{nonrel}) is $\simeq 1.0$. Thus
one may conclude that even for the decays with the smallest pion momentum
$k$, the non-relativistic approximation overestimates the calculated decay
widths by about a factor 2.4.

The calculated decay widths for $D^*\rightarrow D\pi$ as
obtained with $g_A^q=0.87$ and $g_A^q=1$ are given in Table~\ref{pitab}
along with the pion momenta used and the present empirical upper
bounds. The $D$ meson
masses used are those listed in Table~\ref{masstab}. The
calculated total pionic width of the $D^{*\pm}$ is 0.082 MeV in the single
quark approximation, a result, which is about 60\% of the present
empirical upper bound (0.131 MeV). In the single quark approximation the
calculated width of the $D^{*0}$ is 0.036 MeV, which is far below
the empirical upper bound of 1.3 MeV. In this case, the empirical upper
bound is far too large for being theoretically constraining. The 
widths obtained 
here for the $D^*$ mesons are similar to those obtained in 
ref.~\cite{Goity}, where the Gross equation framework was applied. 
In that reference, the value suggested for $g_A^q$ was however only 0.75,
and with that value the present calculated widths would be somewhat
smaller. The difference may be attributed partly to the much
lighter constituent quark masses used in ref.~\cite{Goity} and partly to
the relativistic factors in eq.~(\ref{matrix}), which arise from the
canonical boosts. Without those factors the calculated width of the
$D^{*\pm}$ would exceed the experimental upper bound in the single quark
approximation.

\subsection{Axial exchange current contribution}

The single quark amplitude, eq.~(\ref{oper}), represents a coupling of the
pion to the axial current of the light constituent quark, which
in the static approximation may be expressed as

\begin{equation}
\vec A_a = -g_A^q\vec\sigma_q \tau_a.
\label{confcurr}
\end{equation}
The linear scalar confining interaction will contribute an
exchange current that arises from coupling of the pion
to a virtual negative energy quark, see Fig.~\ref{diagram}. If the
confining interaction is taken to be of the form $V_c(r) = cr - b$, as in
ref.~\cite{Lahde}, then the expression for the corresponding exchange
current operator may, to lowest order in $v/c$, be written
as~\cite{Tsushima}:

\begin{equation}
\vec A_a^{\mathrm{ex}}=-{g_A^q\over 4
m^3}\:\left[\left(cr-b\right)\,\left(3\vec\sigma_q
\vec P^2 -4\vec P\,\vec\sigma_q\cdot\vec P\right)
+{c\over 2r}\vec\sigma_q - {2c\over r}\vec r\times\vec P\right]\:\tau_a.
\label{acurr}
\end{equation}
This expression shows that the exchange current contribution
is a relativistic correction of the same order in $v/c$
as the relativistic corrections to the single-quark operator
in eq.~(\ref{oper}). The constant $b$ was obtained as 320 MeV in
ref.~\cite{Lahde}.

The expression~(\ref{acurr}) does not include the corrections from the
canonical boost factors on the single quark spinors that are
included in the single quark operator, eq.~(\ref{oper}). Hence a more 
realistic evaluation requires that those factors are taken into
account. For simplicity, the same spinor factors
as for the single quark operator are used in this context, and the
operator is also multiplied with the corresponding - numerically less
important - spinor normalization factors for the scalar
vertex on the heavy antiquark line. This implies going beyond
the static local approximation for the confining
potential. Accordingly the potential function $cr$ is to
be replaced with the general expression $c\left|\vec r\,'
+\vec r\,\right|/2$. 

Moreover the factor $1/m^3$ in the axial exchange current
operator~(\ref{confcurr}) arises as the static approximation to a
combination of energy dependent factors appearing in the denominator for
the intermediate negative energy state and of corresponding energy
factors in the quark spinors. Consequently, the static approximation
to these factors implies a very large overestimate of the
axial exchange current contribution. In order to obtain
a more realistic estimate for this contribution, the static approximation
in the axial exchange current operator may be replaced as

\begin{equation}
{1\over m^3}\rightarrow {4\over 2m+E+E'}\left({2m\over E+E'}\right)^2.
\label{repl}
\end{equation}
The resulting matrix element of the exchange current
operator~(\ref{acurr}) between the triplet and singlet states may then be
expressed as

\begin{eqnarray}
{\cal M}_0^{\mathrm{ex}} &=& \frac{5}{12\pi}
\int_{0}^{\infty}dr'r'u_0(r')\int_{0}^{\infty}dr\,r\,
u_0(r)\:V_c\left(\sqrt{\frac{r^{'2}+r^2}{2}}\:\right) 
\int_{0}^{\infty}dP\,P^4 \nonumber \\
&&\int_{-1}^{1}dz\,f_{\mathrm{BS}}(P,z)\:
j_0\left(r'\sqrt{P^2+{k^2\over 16}+{Pkz\over 2}}\:\right)
j_0\left(r\sqrt{P^2+{k^2\over 16}-{Pkz\over 2}}\:\right) \nonumber \\
&&{4\over 2m+E+E'}\left({2\over E+E'}\right)^2
\sqrt{E'+m\over 2 E'}\sqrt{E+m\over 2 E}
\left(1-{P^2-k^2/4\over 3(E'+m)(E+m)}\right) \nonumber \\
&&\sqrt{E_c'+M\over 2 E_c'}\sqrt{E_c+M\over 2 E_c}
\left(1-{P^2-k^2/4\over (E_c'+M)(E_c+M)}\right).
\label{matrix2}
\end{eqnarray}
In the above expression, a symmetrized form $\sqrt{(r^{'2}+r^2)/2}$ of the
confining potential has been employed for simplicity. This form leads to
the correct limit when $r = r'$. Moreover, the
term proportional to $c/2r$ in eq.~(\ref{acurr}) has been neglected in the
calculations because of its smallness. The numerical value of the exchange
current matrix element ${\cal M}_0^{\mathrm{ex}}$ as evaluated using
eq.~(\ref{matrix2}) is obtained as 0.040. In the static limit,
i.e. without the replacement of eq.~(\ref{repl}), that value increases by
about a factor 3. The static limit is thus seen to represent a significant 
overestimate of this effect. In any case, the correction is
non-negligible in comparison with the contribution from the
single quark operator, which is ${\cal M}_0=0.649$. Addition of the
exchange current contribution increases the calculated pionic decay widths
by about 15\%, and brings them to the empirical upper limits in 
the case of the charged $D^*$ mesons, if $g_A^q=1$. These results are also
shown in Table~\ref{pitab}. 

Even though the
total widths of the $D^{*\pm}$ states have not yet been experimentally
determined, their branching fractions for $\pi$ decay are known to a good
degree of accuracy. The current experimental results are $(68.3 \pm
1.4)$\% for $D^{*\pm} \rightarrow D^0 \pi^{\pm}$ and $(30.6 \pm 2.5)$\%
for $D^{*\pm} \rightarrow D^{\pm} \pi^0$ respectively~\cite{PDG}. Using
the results in Table~\ref{pitab}, the ratio of these decay modes is
obtained as
\begin{equation}
\frac{\Gamma \left(D^{*\pm} \rightarrow D^0 \pi^{\pm}\right)}{\Gamma
\left(D^{*\pm} \rightarrow D^{\pm} \pi^0\right)} = 2.2,
\end{equation}
being thus in excellent agreement both with the result obtained in
ref.~\cite{Goity} and the experimentally determined value $2.23 \pm 0.19$.

\newpage

\begin{table}[h!]
\begin{center}
\begin{tabular}{l|c|c|c|c|c}
\quad Decay & $\pi$ momentum $k$ & SQA & SQA + EXCH &$g_A^q=1$&
Experiment\\ \hline\hline  
$D^{*\pm} \rightarrow D^\pm \pi^0$ & 38.3 & 0.026 & 0.029 & 
0.038 & $<0.04$ \\
 &&&& \\
$D^{*\pm} \rightarrow D^0\pi^\pm$ & 39.6 & 0.056 & 0.064 & 
0.084 & $<0.09$ \\
 &&&& \\
$D^{*0}\rightarrow D^0\pi^0$ & 43.1 & 0.036 & 0.041 & 
0.054 & $<1.3$ \\
\end{tabular}  
\caption{The calculated pionic widths and experimental upper limits in
MeV for the $D^*$ mesons, corresponding to $g_A^q$ = 0.87.
The single quark approximation, with relativistic corrections
is denoted SQA, and the result obtained with the exchange current
contribution included is denoted SQA + EXCH. The net calculated widths are
also shown for $g_A^q=1$.}
\label{pitab}
\end{center}
\end{table}

\section{Pion decay widths of the $D_1$ and $D_2^*$ mesons}

\subsection{Pion decay by the single quark axial current}

Only two of the four expected charm meson resonances with $L=1$ have
hitherto been discovered with certainty. These are the $D_1(2420)$ and
the $D_2^*(2460)$ mesons, including all the different charge states. It is
generally assumed that these are spin triplet states with $J=1$ and $J=2$
respectively. In the Heavy Quark Symmetry (HQS) framework, they are
assumed to be states with light quark angular momentum $j_q = 3/2$. For
these $D$ meson states the total widths have been
experimentally determined, although with quite large uncertainty ranges, 
and some of their pionic decays have been "seen"~\cite{PDG}. The basic
pionic decay mode of these resonances is $D-$wave decay by pion coupling
to the axial current operator, eq.~(\ref{oper}), of the charm mesons. In
the case of the $D_1$ meson, $S-$wave pion decay through the axial
charge operator also contributes significantly to the decay
width~\cite{Mehen}.

The axial current transition operator for pion decay to the ground state
$D$ mesons is given by eq.~(\ref{oper}). If both the $D_1(2420)$ and 
$D_2^*(2460)$ mesons are assumed to be mainly spin triplet
states, the calculation of the pion decay widths of these states require
the following spin sums:

\begin{eqnarray}
S_s&=&{1\over 2J+1}\sum_{M=-J}^{J}\left<11,JM\right|\vec \sigma_q\cdot \vec
k\left|00,00\right>\left<00,00
\right|\vec \sigma_q\cdot \vec k\left|11,JM\right>
\nonumber \\
&=&{1\over 2J+1}\sum_{M=-J}^{J}\left<11,JM\right|{1\over 3}k^2-{1\over
6}S_{12}(\vec k)\left|11,JM\right>, \label{ss1}
\end{eqnarray}
for spin singlet final states, and

\begin{eqnarray}
S_t&=&{1\over 2J+1}\sum_{M=-J}^{J}\sum_{m=-1}^{1}
\left<11,JM\right|\vec \sigma_q
\cdot \vec k\left|01,1m\right>\left<01,1m\right|
\vec \sigma_q\cdot \vec k\left|11,JM\right>
\nonumber \\
&=&{1\over 2J+1}\sum_{M=-J}^{J}\left<11,JM\right|{2\over 3}k^2+{1\over
6}S_{12}(\vec k)\left|11,JM\right>, \label{ss2}
\end{eqnarray}
for spin triplet final states. In the above expressions, $S_{12}$ denotes
the tensor operator

\begin{equation}
S_{12}(\hat k) = 3\vec\sigma_q\cdot \hat k\,\vec\sigma_{\bar
Q}\cdot \hat k -\vec\sigma_q\cdot\vec\sigma_{\bar Q}.
\end{equation}
Evaluation of the
orbital matrix elements of these operators then leads to the following
expressions for pion decay driven by the axial current operator:

\begin{eqnarray}
\Gamma_A\left(D_1 \rightarrow D\pi\right)&=&0,
\label{forb}\\
\Gamma_A\left(D_1 \rightarrow D^*\pi\right)&=&{3\over
8\pi}{E_{D^*}\over M_{D_1}}
\left({g_A^q\over f_\pi}\right)^2k^3{\cal M}_1^2, \label{dec1}
\end{eqnarray}
for the $D_1$ meson, and

\begin{eqnarray}
\Gamma_A\left(D_2^*\rightarrow D\pi\right)&=&{3\over 8\pi}{2\over 5}
{E_D\over M_{D_2^*}}
\left({g_A^q\over f_\pi}\right)^2k^3{\cal M}_1^2,\label{dec2}\\
\Gamma_A\left(D^*_2\rightarrow D^*\pi\right)&=&{3\over 8\pi}{3\over 5}
{E_{D^*}\over M_{D_2^*}}\left({g_A^q\over
f_\pi}\right)^2k^3{\cal M}_1^2.\label{dec3}
\end{eqnarray}
for the $D_2^*$ meson. In the above expressions, all final charge states
have been included. ${\cal M}_1$ is an orbital matrix element defined in
analogy with eq.~(\ref{matrix}) and may be expressed as 

\begin{eqnarray}
{\cal M}_1&=&{1\over \pi}\int_{0}^{\infty}dr'r'u_0(r')\int_{0}^{\infty}dr\,r\,
u_1(r)\int_{0}^{\infty}dP\,P^2\int_{-1}^{1}dz\,f_{\mathrm{BS}}(P,z) 
\nonumber \\
&&\sqrt{{E'+m\over 2E'}}\sqrt{{E+m\over 2E}}\left(1-{P^2-k^2/4\over 3
(E'+m)(E+m)}\right) 
{k/4-Pz\over \sqrt{P^2+k^2/16-Pkz/2}}
\nonumber \\
&&j_0\left(r'\sqrt{P^2+{k^2\over 16}+{Pkz\over
2}}\:\right)j_1\left(r\sqrt{P^2+{k^2\over
16}-{Pkz\over 2}}\:\right). \label{pmatr}
\end{eqnarray}
In eq.~(\ref{pmatr}), the factor $f_{\mathrm{BS}}(P,z)$ is defined as in
eq.~(\ref{bfact}), and the quark energy factors $E,E'$ are defined as for 
eq.~(\ref{matrix}). $u_1(r)$ is the reduced radial wave function for the
$D$ meson $P$-state

\begin{equation}
\left|11,JM\right>=\sum_{ls}\left<11 ls|JM\right|{u_1(r)\over
r}Y_{1l}(\hat r)\left|1s\right>,
\end{equation}
where $\left|1s\right>$ denotes a spin triplet state with $s_z=s$. In the
non-relativistic limit the matrix element~(\ref{pmatr}) reduces to

\begin{equation}
{\cal M}_1=\int_{0}^{\infty}dr\,u_0(r)\,u_1(r)\,j_1\left({kr\over
2}\right).
\label{nonrel2}
\end{equation}

Note that eq.~(\ref{forb}) predicts that the $D_1$ meson cannot
decay to $D\pi$. This prediction only holds if the $D_1$ is a state with
$J=1$. As will be shown, it does not provide a means to discriminate
between spin singlet and triplet states. However, the prediction that the
$D_2^*$ decays to both $D^*$ and $D$ permits the identification of the
empirical $D_2^*(2460)$ meson as a spin triplet state with $J=2$. The
numerical values for the matrix element ${\cal M}_1$, as obtained with
the $L=0$ and $L=1$ wave functions shown in
Fig.~\ref{wavefig} for the pionic decay modes~(\ref{dec1}-\ref{dec3}), are
listed in Table~\ref{tab2}. In all cases, the non-relativistic
approximation overestimates the calculated matrix elements by $\sim50\%$,
which is to be expected in view of the relatively small masses of the
light constituent quarks. The calculated pionic decay widths corresponding
to the matrix elements in Table~\ref{tab2} are given in Table~\ref{tab3}
along with the current empirical values.

\subsection{Axial exchange current contribution}

The axial exchange current operator, eq.~(\ref{acurr}), may also
contribute significantly to the~pionic decay widths of the $L=1$ charm
mesons. The orbital matrix element of the spin part of the exchange
current operator between the $L=1$ and $L=0$ states, when evaluated with
the same spinor factors and approximations as in eq.~(\ref{matrix2}), may
be expressed as

\begin{eqnarray}
{\cal M}_1^{\mathrm{ex}}&=&\frac{5}{12\pi}
\int_{0}^{\infty}dr'r'u_0(r')\int_{0}^{\infty}dr\,r\,u_1(r)\:
V_c\left(\sqrt{\frac{r^{'2}+r^2}{2}}\:\right) \int_{0}^{\infty}dP\,P^4 
\int_{-1}^{1}dz\,f_{\mathrm{BS}}(P,z) \nonumber \\
&&j_0\left(r'\sqrt{P^2+{k^2\over 16}+{Pkz\over 2}}\:\right)
j_1\left(r\sqrt{P^2+{k^2\over 16}-{Pkz\over 2}}\:\right) 
{k/4-Pz\over \sqrt{P^2+k^2/16-Pkz/2}} \nonumber \\
&&{4\over 2m+E+E'}\left({2\over E+E'}\right)^2
\sqrt{E'+m\over 2 E'}\sqrt{E+m\over 2 E}
\left(1-{P^2-k^2/4\over 3(E'+m)(E+m)}\right) \nonumber \\
&&\sqrt{E_c'+M\over 2 E_c'}\sqrt{E_c+M\over 2 E_c}
\left(1-{P^2-k^2/4\over (E_c'+M)(E_c+M)}\right).
\label{matrix3}
\end{eqnarray}
The numerical values of these matrix elements are
also listed in Table~\ref{tab2}. Their magnitude corresponds to
$\sim$ 10\% of the
matrix elements of the corresponding single quark operator,
eq.~(\ref{pmatr}), and evidently the exchange current contribution is
numerically significant these cases as well. This contribution
enhances the net calculated pionic decay widths of the $L=1$ charm mesons
and brings them closer to the empirical values. These results are shown
in Table~\ref{tab3}.

\vspace{0.3cm}

\begin{table}[h!]
\begin{center}
\begin{tabular}{l|c|c|c}
\quad Decay & SQA-NR & SQA-REL & EXCH \\ \hline\hline
$D_1\rightarrow D^*\pi$   & 0.135 & 0.093 & 0.010 \\ && \\
$D_2^*\rightarrow D\pi$   & 0.185 & 0.126 & 0.012 \\ && \\
$D_2^*\rightarrow D^*\pi$ & 0.147 & 0.101 & 0.011 \\ 
\end{tabular}
\caption{Matrix elements ${\cal M}_1$ of the single quark axial current
operator in the non-relativistic approximation, eq.~(\ref{nonrel2}), and
without approximation, eq.~(\ref{pmatr}), for pion decay of the $D_1$ and
$D_2^*$ mesons. The column ``exchange'' contains the matrix
elements ${\cal M}_1^{\mathrm{ex}}$ corresponding to 
eq.~(\ref{matrix3}) of the spin part of the axial exchange current
operator~(\ref{acurr}).}
\label{tab2}
\end{center}
\end{table}

\subsection{Pion decay by the axial charge}

The charge component of the axial vector coupling in eq.~(\ref{lagr}) also
contributes to the decays of the $D$-mesons with $L=1$, and gives rise to
both $S$- and $D$-wave pion decay. The amplitude describing the coupling
of the pion field to the axial charge component of the light constituent
quark may then be obtained from eq.~(\ref{lagr}) as

\begin{equation}
T_S=i{g^q_A\over 2f_\pi}{2m+E+E'\over
\sqrt{4EE'(E+m)(E'+m)}}\:\omega_\pi\,\vec \sigma\cdot \left({\vec p\,'
+\vec p\over 2}\right)\,\tau_\pi,
\label{onebody}
\end{equation}
where $\omega_\pi=\sqrt{k^2+m_\pi^2}$ is the energy of the emitted
pion. In the evaluation of the matrix element of $T_S$ between $D$-meson
states with $L=1$ and the ground state charm mesons, the operator
$\left(\vec p\,'+\vec p\,\right)/2$ is treated as the differential
operator $i\,(\vec \nabla\,' - \vec \nabla)/2$. Application of the
operators on the initial and final orbital wave functions leads to spin
matrix elements of the form

\begin{equation}
\left<S'M\right|\sum_{ls}\left<11ls|JM\right>\,\sigma_l^1\,\left|1s\right> 
=-\delta_{JS'}\delta_{M'M}\left[\sqrt{3}\delta_{S'0}+
\sqrt{2}\delta_{S'1}\right],
\end{equation}
from which it follows that $S$-wave pion decay can only contribute to the
$D_1\rightarrow D^*\pi$ decay widths. The resulting contribution to the 
$D_1(2420)\rightarrow D^*\pi$ decay width from $S$-wave pion decay may be
written in the form

\begin{equation}
\Gamma_C\,\left(D_1\rightarrow D^*\pi\right)={1\over
32\pi}{E_{D^*}\over M_{D_1}}\left({g^q_A\over
f_\pi}\right)^2\left({\omega_\pi\over m}\right)^2\,k\,
{\cal M}_{1\mathrm S}^2.
\label{s-decp}
\end{equation}
In eq.~(\ref{s-decp}), the matrix element ${\cal M}_{1\mathrm S}$ 
is defined, in analogy with the matrix
elements in eqs.~(\ref{matrix}) and~(\ref{pmatr}), as

\begin{eqnarray}
{\cal M}_{1\mathrm S}&=&{1\over
\pi}\int_{0}^{\infty}dr'r'\int_{0}^{\infty}dr\,r\int_{0}^{\infty}
dP\,P^2 \int_{-1}^{1}dz\,f_{\mathrm{BS}}(P,z) \nonumber \\
&&{m(2m+E+E')\over \sqrt{4EE'(E+m)(E'+m)}}
\left[u'_0(r')u_1(r)-u_0(r')u'_1(r)-2{u_0(r')u_1(r)\over r}\right]
\nonumber \\
&&j_0\left(r'\sqrt{P^2+{k^2\over
16}+{Pkz\over 2}}\:\right)j_0\left(r\sqrt{P^2+{k^2\over 16}-{Pkz\over
2}}\:\right).
\label{1Selem}
\end{eqnarray}
In the above expression, the factors $u'(r)$ denote the derivatives of the
reduced radial wavefunctions displayed in Fig.~\ref{wavefig}. In the
non-relativistic limit, the matrix element~(\ref{1Selem}) simplifies to

\begin{equation}
{\cal M}_{1\mathrm
S}=\int_{0}^{\infty}dr\left[u'_0(r)u_1(r)-u_0(r)u'_1(r)-2{u_0(r)u_1(r)\over
 r}\right]\,j_0\left({kr\over 2}\right).
\end{equation}
The numerical values of these matrix elements are given in
Table~\ref{tab4}. The values of the matrix elements 
${\cal M}_{1\mathrm S}$ in Table~\ref{tab4} are very large and lead to
unrealistically large contributions from $S$-wave pion decay to the widths
of the $D_1$ mesons. Insertion in eq.~(\ref{s-decp}) would give for
$\Gamma_C\left(D_1\rightarrow D^*\pi\right)$ the value
$\sim 60$ MeV, which exceeds the empirically determined width by about a
factor 4. Reduction of this value to a realistic level is however brought
about by the two-body mechanism described in Section~\ref{conf_sec} that
is implied by the linear confining interaction, if its spinor structure is
scalar. 

The amplitude~(\ref{onebody}) also gives rise to $D$-wave pion decay,
which contributes to the decay widths of both the $D_1$ and
$D_2^*$ mesons. The explicit expressions for these contributions from
$D$-wave pion decay may be obtained as

\begin{equation}
\Gamma_D\left(D_1\rightarrow D^*\pi\right)=\frac{1}{8\pi}
\frac{E_{D^*}}{M_{D_1}}
\left(\frac{g_A^q}{f_{\pi}}\right)^2
\left(\frac{\omega_{\pi}}{m}\right)^2\,k\,{\cal M}_{1\mathrm D}^2
\label{tiny1}
\end{equation}
for the $D_1$ meson, and

\begin{eqnarray}
\Gamma_D\left(D_2^*\rightarrow D\pi\right)&=&\frac{3}{8\pi}\frac{2}{5}
\frac{E_D}{M_{D_2^*}} \left(\frac{g_A^q}{f_{\pi}}\right)^2
\left(\frac{\omega_{\pi}}{m}\right)^2\,k\,{\cal M}_{1\mathrm D}^2, \\
\Gamma_D\left(D_2^*\rightarrow D^*\pi\right)&=&\frac{3}{8\pi}\frac{3}{5}
\frac{E_{D^*}}{M_{D_2^*}} \left(\frac{g_A^q}{f_{\pi}}\right)^2
\left(\frac{\omega_{\pi}}{m}\right)^2\,k\,{\cal M}_{1\mathrm D}^2.
\label{tiny3}
\end{eqnarray}
for the $D_2^*$ meson. In the above expressions, the matrix element
${\cal M}_{1\mathrm D}$ is again defined analogously to
eq.~(\ref{1Selem}), and may be expressed as

\begin{eqnarray}
{\cal M}_{1\mathrm D}&=&-{1\over
\pi}\int_{0}^{\infty}dr'r'\int_{0}^{\infty}dr\,r\int_{0}^{\infty}
dP\,P^2 \int_{-1}^{1}dz\,f_{\mathrm{BS}}(P,z)\,
\left(\frac{3}{2}\frac{P^2(1-z^2)}{P^2+{k^2\over 16}-{Pkz\over
2}}-1\right)
\nonumber \\
&&{m(2m+E+E')\over \sqrt{4EE'(E+m)(E'+m)}}
\left[u'_0(r')u_1(r)-u_0(r')u'_1(r)+{u_0(r')u_1(r)\over r}\right]
\nonumber \\
&&j_0\left(r'\sqrt{P^2+{k^2\over
16}+{Pkz\over 2}}\:\right)j_2\left(r\sqrt{P^2+{k^2\over 16}-{Pkz\over
2}}\:\right).
\label{1Delem}
\end{eqnarray}
In the non-relativistic limit, this expression reduces to

\begin{equation}
{\cal M}_{1\mathrm
D}\simeq\int_{0}^{\infty}dr\left[u'_0(r)u_1(r)-u_0(r)u'_1(r)
+{u_0(r)u_1(r)\over  r}\right]\,j_2\left({kr\over 2}\right).
\label{tinyeq}
\end{equation}
The numerical values of these matrix elements for $D-$wave decay as
obtained with the wavefunctions of ref.~\cite{Lahde} turn out to be
exceedingly tiny in comparison with those from $S$-wave decay. The reason
for this is immediately apparent if one considers the wave function
combination $u'_0 u_1 - u_0 u'_1 + u_0 u_1/r$ that appears in the
integrand in eq.~(\ref{tinyeq}). If the approximate wave functions plotted
in Fig.~\ref{wavefig} and given in Appendix~A are used, that wave function
combination vanishes exactly. As these approximate wavefunctions are not
perfect and as the nonlocal structure of the integrand in 
eq.~(\ref{1Delem}) has to be taken into account, this cancellation is not 
absolute if the unapproximated expressions and wave functions are used.
The numerical values of eqs.~(\ref{1Delem}) and~(\ref{tinyeq}) turn
out to be of the order 1 MeV and thus completely negligible as compared to
the matrix elements of $S-$wave decay listed in Table~\ref{tab4}. These
$D$-wave decay amplitudes that arise from the axial charge operator also
obtain a contribution from the two-body mechanism outlined in
Section~\ref{conf_sec}.

\newpage

\subsection{Two-quark contributions to the axial charge operator}
\label{conf_sec}

Both the confining and one-gluon exchange~(OGE) interactions contribute
two-body terms to the axial charge density operator.
If the linear confining interaction has the form $V_c(r)\:\mathcal S$,
where $V_c$ is the central confining potential, and~$\mathcal S$ is the
scalar (Fermi) invariant for the two-quark system, it gives rise to an
axial exchange charge density operator, which to lowest order in $v/c$
may be expressed as~\cite{Mariana}
\begin{equation}
A_{\mathrm{conf}}^0={g_A^q\over m^2}\,V_c(r)\,\vec \sigma\cdot \vec P\,
\tau_a.
\end{equation} 
The corresponding expression for the vector coupled one-gluon
exchange interaction is
\begin{equation}
A_{\mathrm{OGE}}^0={g_A^q\over m M}\,V_g(r)\,
\vec \sigma\cdot\vec P_{\bar c}\, \tau_a,
\label{glu}
\end{equation} 
where the momentum operator $\vec P_{\bar c}$ is defined as 
$\vec P_{\bar c}=(\vec p_{\bar c}\,' +\vec p_{\bar c})/2.$
Here $\vec p_{\bar c}$ and $\vec p_{\bar c}\,'$ denote the momentum
operators of the heavy quark and $V_g(r)$ is the main
spin-independent term in the one-gluon exchange potential. If the color
couplant $\alpha_s$ is taken to be constant, then $V_g(r)$ may be 
expressed as $V_g(r)=-4\alpha_s/3r$ in the static limit.

\begin{figure}[h!]
\begin{center}
\begin{tabular}{c c c} \\
\begin{fmffile}{Zgraph}
\begin{fmfgraph*}(130,150) \fmfpen{thin}
\fmfleft{i2}
\fmftop{i3,o2}
\fmfbottom{i1,o1}
\fmf{dashes,label=$\pi$}{i2,v1}
\fmf{fermion}{i1,v2,v1,i3}
\fmf{dashes,label=$V_c$}{v2,v3}
\fmf{fermion}{o2,v3,o1}  
\fmflabel{$q$}{i1}
\fmfforce{(.25w,.35h)}{v1}
\fmfforce{(.45w,.5h)}{v2}
\fmfforce{(.9w,.5h)}{v3}
\fmfforce{(.35w,0)}{i1}
\fmfforce{(0,.35h)}{i2}
\fmfforce{(.15w,h)}{i3}
\fmfforce{(w,0)}{o1}
\fmfv{l=$\bar Q$,l.a=-90,l.d=.04w}{o1}
\fmfforce{(w,h)}{o2}
\end{fmfgraph*}
\end{fmffile}
& \quad \quad \quad \quad &
\begin{fmffile}{Zgraph2} 
\begin{fmfgraph*}(130,150) \fmfpen{thin}
\fmfleft{i2}
\fmftop{i3,o2}
\fmfbottom{i1,o1}   
\fmf{dashes,label=$\pi$}{i2,v1}
\fmf{fermion}{i1,v1,v2,i3}
\fmf{dashes,label=$V_c$}{v2,v3}
\fmf{fermion}{o2,v3,o1}
\fmflabel{$q$}{i1}
\fmfforce{(.25w,.65h)}{v1}
\fmfforce{(.45w,.5h)}{v2}
\fmfforce{(.9w,.5h)}{v3}
\fmfforce{(.15w,0)}{i1}
\fmfforce{(0,.65h)}{i2}  
\fmfforce{(.35w,h)}{i3}
\fmfforce{(w,0)}{o1}
\fmfv{l=$\bar Q$,l.a=-90,l.d=.04w}{o1}
\fmfforce{(w,h)}{o2}
\end{fmfgraph*}
\end{fmffile}
\\
\\
\\
\end{tabular}
\caption{Two-quark contributions to the pion production amplitude
associated with the axial charge operator. The diagrams shown correspond
to both time orderings of the two-quark contribution from the scalar
confining interaction. Two similar diagrams arise from the one-gluon
exchange interaction, in which case the scalar vertices are to be
appropriately replaced.} \label{diagram} 
\end{center}
\end{figure}
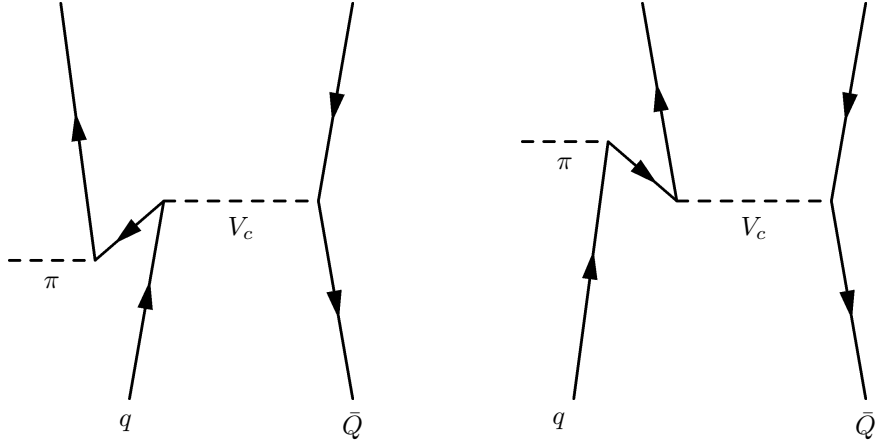

In the static limit, the two-body contribution that arises from the 
confining and one-gluon exchange interactions to the amplitude for
$S$-wave pion decay may be expressed in the form

\begin{equation}
T_S^{(2)} = - \frac{1}{m}\left(cr -b
+\frac{4}{3}\frac{m}{M}\frac{\alpha_s}{r}\right) T_S.
\label{twobody}
\end{equation}
Here, $T_S$ denotes the single quark amplitude for $S-$wave pion
decay, eq.~(\ref{onebody}). This result may be derived as a pair current
operator, or more simply by making the scalar shift in the light
constituent quark mass $m \rightarrow m + V_c(r)$ in the case of the
scalar confining interaction. This two-body operator arises in the
non-relativistic reduction of the general pion decay amplitude for the
$q\bar Q$ system as a pair (or seagull) term, see Fig.~\ref{diagram}. A
corresponding two-body term appears in the amplitude for pion production
in nucleon-nucleon collisions, the difference being that $V_c$ is in that
case replaced by the effective scalar component of the nucleon-nucleon
interaction. 

In eq.~(\ref{twobody}) the factor $1/m$ represents the static limit of the
propagator of the intermediate negative energy quark. As the
static limit is known, from the treatment of the axial exchange current, 
to give large overestimates in the case of light
constituent quarks, the static propagator will here be replaced by the
symmetrized form $4/(2m+E+E')$. The matrix element of the two-quark
contribution to the $S-$wave decays through the axial charge operator may
then be obtained by modifying the single quark matrix element for $S$-wave
pion decay, eq.~(\ref{1Selem}), according to eq.~(\ref{twobody}), giving

\begin{eqnarray}
{\cal M}_{2\mathrm S}^{\mathrm{conf}} &\simeq& -{1\over \pi}
\int_{0}^{\infty}dr'r'\int_{0}^{\infty}dr\,r\:
V_c\left(\sqrt{\frac{r^2 + r'^2}{2}}\:\right)\,
\int_{0}^{\infty}dP\,P^2 \int_{-1}^{1}dz\,f_{\mathrm{BS}}(P,z)
\nonumber \\
&&{4m\over \sqrt{4EE'(E+m)(E'+m)}}
\left[u'_0(r')u_1(r)-u_0(r')u'_1(r)-2{u_0(r')u_1(r)\over r}\right]
\nonumber \\
&&j_0\left(r'\sqrt{P^2+{k^2\over
16}+{Pkz\over 2}}\:\right)j_0\left(r\sqrt{P^2+{k^2\over 16}-{Pkz\over
2}}\:\right) \nonumber \\
&&\sqrt{E_c'+M\over 2 E_c'}\sqrt{E_c+M\over 2 E_c}
\left(1-{P^2-k^2/4\over (E_c'+M)(E_c+M)}\right).
\label{2Selem}
\end{eqnarray}
In the above expression, the symbol $\simeq$ indicates that $|\vec r +
\vec r\,'|/2$ has been approximated by the expression
$\sqrt{(r^{'2}+r^2)/2}$ as in eq.~(\ref{matrix2}). In the nonrelativistic
limit, eq.~(\ref{2Selem}) becomes

\begin{equation}
{\cal M}_{2\mathrm S}^{\mathrm{conf}} \simeq 
-\frac{1}{m}\int_{0}^{\infty}dr\,V_c(r)\,\left[u'_0(r)u_1(r)-u_0(r)u'_1(r)
-2{u_0(r)u_1(r)\over r}\right]\,j_0\left({kr\over 2}\right).
\end{equation}

The gluon exchange contribution, eq.~(\ref{glu}), to the axial exchange
charge operator gives a contribution of smaller magnitude than the
confining interaction because of the heavy antiquark mass involved. Since
the heavy quark constituent mass, 1580 MeV, is not very large compared to
the light quark mass of 450 MeV, the one-gluon exchange contribution turns
out to be significant as well, see Table~\ref{tab4}, and cannot be
neglected. It contributes the following orbital matrix element
for pion decay of the $D_1$ meson:
 
\begin{eqnarray}
{\cal M}_{2\mathrm S}^{\mathrm{OGE}} &\simeq& {1\over \pi}
\int_{0}^{\infty}dr'r'\int_{0}^{\infty}dr\,r\,
V_g\left(\sqrt{\frac{r^2 + r'^2}{2}}\,\right)\,
\int_{0}^{\infty}dP\,P^2 \int_{-1}^{1}dz\,f_{\mathrm{BS}}(P,z)
\nonumber \\
&&{4m^2 \over \sqrt{4EE'(E+m)(E'+m)}}\quad
{2M+E_c'+E_c \over \sqrt{4 E_c' E_c(E_c'+M)(E_c+M)}}
\nonumber \\
&&\left[u'_0(r')u_1(r)-u_0(r')u'_1(r)-2{u_0(r')u_1(r)\over r}\right]
\nonumber \\
&&j_0\left(r'\sqrt{P^2+{k^2\over
16}+{Pkz\over 2}}\:\right)j_0\left(r\sqrt{P^2+{k^2\over 16}-{Pkz\over
2}}\:\right).
\label{2SelemG}
\end{eqnarray}
The effective one-gluon exchange potential $V_g(r)$ that is
appropriate for the $D$ meson systems and which takes
into account the screened running color coupling $\alpha_s$
has been calculated in ref.~\cite{Lahde}. This potential function
may be well parametrized as $V_g(r)=-A\, \arctan(Br)/r$,
with A=1.1899 and B=3.39768/fm. Using this form, the numerical value 
${\cal M}_{2\mathrm S}^{\mathrm{OGE}}$= 132 MeV is obtained. In the static 
non-relativistic limit, the one-gluon exchange contribution reduces to

\begin{equation}
{\cal M}_{2\mathrm S}^{\mathrm{OGE}} \simeq 
\frac{1}{M}\int_{0}^{\infty}dr\,V_g(r)\,\left[u'_0(r)u_1(r)-u_0(r)u'_1(r)
-2{u_0(r)u_1(r)\over r}\right]\,j_0\left({kr\over 2}\right).
\end{equation}

The two-quark mechanism shown in Fig.~\ref{diagram} contributes to the
$D$-wave pion decay amplitude from the axial charge operator as well. As
in the case of $S$-wave pion decay, this effect may be taken into account
by adding to the one-body matrix element ${\cal M}_{1\mathrm D}$ the 
matrix element that is associated with the exchange charge
contribution. For the confining interaction, this matrix element may be
obtained similarly to eq.~(\ref{2Selem}), giving

\begin{eqnarray}
{\cal M}_{2\mathrm D}^{\mathrm{conf}} &\simeq& {1\over\pi}
\int_{0}^{\infty}dr'r'\int_{0}^{\infty}dr\,r\,
V_c\left(\sqrt{\frac{r^2 + r'^2}{2}}\:\right) 
\int_{0}^{\infty}dP\,P^2 \int_{-1}^{1}dz\,f_{\mathrm{BS}}(P,z)
\nonumber \\
&&\left(\frac{3}{2}\frac{P^2(1-z^2)}{P^2+{k^2\over 16}-{Pkz\over
2}}-1\right)\:\left[u'_0(r')u_1(r)-u_0(r')u'_1(r)+{u_0(r')u_1(r)\over
r}\right]
\nonumber \\
&&{4m \over \sqrt{4EE'(E+m)(E'+m)}}\:
\sqrt{E_c'+M\over 2 E_c'}\sqrt{E_c+M\over 2 E_c}
\left(1-{P^2-k^2/4\over (E_c'+M)(E_c+M)}\right)
\nonumber \\
&&j_0\left(r'\sqrt{P^2+{k^2\over
16}+{Pkz\over 2}}\:\right)j_2\left(r\sqrt{P^2+{k^2\over 16}-{Pkz\over
2}}\:\right).
\label{2Delem}
\end{eqnarray}
A similar expression may readily be constructed for the $D-$wave
contribution from the one-gluon exchange interaction by the techniques
outlined in this section. The numerical value of the matrix
element~(\ref{2Delem}) is of the same magnitude (1 MeV) as that of
eq.~(\ref{1Delem}) and thus completely insignificant. As the one-gluon
exchange contribution is smaller by a factor $m/M$, it will not be
considered in this paper. In the nonrelativistic limit, the
expression~(\ref{2Delem}) reduces to

\begin{equation}
{\cal M}_{2\mathrm D}^{\mathrm{conf}}\simeq 
-\frac{1}{m}\int_{0}^{\infty}dr\,V_c(r)\,\left[u'_0(r)u_1(r)-u_0(r)u'_1(r)
+{u_0(r)u_1(r)\over r}\right]\,j_2\left({kr\over 2}\right).
\end{equation}
The numerical values of the matrix elements for $S-$wave pion decay of the
$D_1$ meson are given in Table~\ref{tab4}. The matrix element of the
two-quark contribution ${\cal M}_{2\mathrm S}^{\mathrm{conf}}$,
eq.~(\ref{2Selem}), is of the same magnitude as the single quark
contribution, eq.~(\ref{1Selem}), and has opposite sign. The corresponding
matrix element from the one-gluon exchange interaction, 
eq.~(\ref{2SelemG}), represents a somewhat smaller contribution. Addition
of these three matrix elements reduces the net matrix element for $S$-wave
pion decay of the $D_1$ state by about a factor~3 as shown in 
Table~\ref{tab4}. With this smaller value the net contribution of $S$-wave
pion decay to $\left(D_1\rightarrow D^*\pi\right)$ becomes quite small and
comparable to that from the axial current operator, as indicated in
Table~\ref{tab3}. It should, however, be noted that the results are very
sensitive to the exact form and composition of the interaction Hamiltonian
used in the calculations.

In the case of the $D_2^*$ mesons the calculated width falls below the
current empirical range by about 40\% if $g_A^q=0.87$. With $g_A^q=1$ the
calculated values are only slightly below the empirical values. These
results leave room for about 5-10 MeV for the contribution from $\pi\pi$
decay. This is what would be expected on the basis of analogy with the
decay pattern of the strange $K_2^*(1430)$ meson, which should have a
structure similar to that of the $D_2^*(2460)$, once the charm quark is
replaced by a strange quark. If the nonrelativistic values for the
orbital matrix elements ${\cal M}_1$ given in Table~\ref{tab2} are
used in the calculation, the calculated width for the $D_2^*$ meson
would be well above the empirical range. The ratio of the 
calculated pion decay widths of the $D_2^*$ meson and the $D_1$ meson
is 1.2. This result is compatible with the current empirical value $\sim
1.3$.

\begin{table}[h!]
\begin{center}
\begin{tabular}{l|r|c|c|c}
\quad\quad Decay & ${\cal M}_{1\mathrm S}$ & ${\cal M}_{2\mathrm
S}^{\mathrm{conf}}$ & ${\cal M}_{2\mathrm S}^{\mathrm{OGE}}$ & Total
\\ \hline\hline 
$D_1\rightarrow D^*\pi$, NR    & -1194 & +450 & +411 & -333  \\
&&& \\
$D_1\rightarrow D^*\pi$, REL   & -592  & +278  & +133 & -181  \\
\end{tabular}
\caption{Matrix elements in MeV of the single quark and two-quark
operators for the $S-$wave axial charge contribution to the decay
$D_1\rightarrow D\pi$. The resulting net contribution to this decay mode
is also given. The labels NR and REL indicate that the non-relativistic
and relativistic expressions have been used respectively.}
\label{tab4}
\end{center}
\end{table}

\begin{table}[h!]
\begin{center}
\begin{tabular}{l|c|c|c|c|c}
\quad\quad Decay & Current & Charge & Total & $g_A^q=1$ 
&Experiment\\ \hline\hline
$D_1\rightarrow D^*\pi$ & 4.2 (3.4) MeV & 6.1 MeV & 10.3 MeV & 13.6 MeV &
$18.9_{-3.5}^{+4.6}$ MeV \\ &&&&& \\
$D_2^*\rightarrow D\pi$ & 8.1 (6.7) MeV & -- & 8.1 MeV & 10.6 MeV
& ? \\ &&&& &\\
$D_2^*\rightarrow D^*\pi$ & 3.9 (3.1) MeV & -- & 3.9 MeV & 5.1 MeV 
&? \\ &&&& \\
$D_2^*\rightarrow D\pi +D^*\pi$ & 11.9 (9.9) MeV & -- & 11.9 MeV &
15.7 MeV & 25$^{+8}_{-7}$ MeV \\
\end{tabular}
\caption{Calculated and empirical pion decay widths of the $D_1$ and
$D_2^*$ mesons driven by the axial current and charge
operators respectively, for $g_A^q = 0.87$. The empirical values are 
total widths~\cite{PDG}, which should mainly be due to pion decay to the
ground state. The numbers in parentheses are the decay widths obtained
without the axial exchange current contribution. The calculated values are
also shown for $g_A^q = 1$.}
\label{tab3}
\end{center}
\end{table}

In the case of the $D_1$ mesons the total width also obtains
a significant contribution from $S$-wave pion decay. Here it has been
assumed that the empirical widths are mainly due to pion decay to the
ground state $D$ and $D^*$ mesons. In addition, even though the present
empirical data on the $L=1$ $D$ mesons is severely limited, the ratio of
$D\pi$ to $D^*\pi$ decay of the $D_2^*$ meson has been
measured~\cite{PDG}. From the results in Table~\ref{tab3}, we obtain

\begin{equation}
\frac{\Gamma \left(D_2^* \rightarrow D\pi\right)}
{\Gamma \left(D_2^* \rightarrow D^*\pi\right)} = 2.1,
\label{ratio}
\end{equation}
which is in agreement with the experimentally determined value
$2.3 \pm 0.6$ for the neutral $D_2^*$ meson~\cite{PDG} (The measured value
for the charged $D_2^*$ meson has much larger statistical
uncertainties). Note that in Table~\ref{tab3}, the quoted experimental
values are for the neutral $D_1$ and the charged $D_2^*$ mesons. The
corresponding value for the neutral $D_2^*$ meson is $23\pm 5$
MeV~\cite{PDG}. The available data for the charged $D_1^*$ meson is
considerably poorer, since it has only recently been discovered.

\vspace{0.5cm}

\section{Pion decay widths of the $D$ mesons with $L=1,J=0,1$}

Two $D$ meson resonances with $L=1$ remain to be found
experimentally. In the $LS$ basis these are the spin singlet state $D_1^*$ 
with $J=1$ and the spin triplet state $D_0^*$ with $J=0$. In this
context, it is worth noting that although the 'star' in the labeling
notation $D^*$ is usually reserved for states with $J^P = 
0^+,1^-,2^+,\cdots$, it has become conventional to label the spin singlet
state $D_1^*$ to distinguish it from the spin triplet state
$D_1$. In the Heavy Quark Symmetry
(HQS) framework, the $D_1^*$ and $D_0^*$ mesons correspond to states with
light quark angular momentum $j_q = 1/2$. Both of these states may pion
decay by $D-$wave decay through the axial current operator as well as 
by $S-$wave decay through the axial charge operator. Consider first the
(mainly) singlet $D_1^*$ state. Because of the
spin dependence of the pion-quark coupling, it follows that this state can
only pion decay to the triplet $D^*$ meson. To derive the
expression for the width of this decay mode, the following additional spin
sums are required:

\begin{eqnarray}
S_s&=&{1\over 3}\sum_{M=-1}^{1}\left<10,1M\right|\vec \sigma_q\cdot \vec
k\left|00,00\right>\left<00,00
\right|\vec \sigma_q\cdot \vec k\left|10,1M\right>
\nonumber \\
&=&{1\over 3}\sum_{M=-1}^{1}\left<10,1M\right|-{1\over
6}S_{12}(\vec k)\left|10,1M\right>,
\end{eqnarray}
for spin singlet final states, and

\begin{eqnarray}
S_t&=&{1\over 3}\sum_{M=-1}^{1}\sum_{m=-1}^{1}
\left<10,1M\right|\vec \sigma_q
\cdot \vec k\left|01,1m\right>\left<01,1m\right|
\vec \sigma_q\cdot \vec k\left|10,1M\right>
\nonumber \\
&=&{1\over 3}\sum_{M=-1}^{1}\left<10,1M\right|k^2+{1\over
6}S_{12}(\vec k)\left|10,1M\right>,
\end{eqnarray}
for spin triplet final states. From these expressions, it follows that the
decay mode $D_1^*\rightarrow D\pi$ is forbidden, since the tensor operator
$S_{12}$ has a vanishing matrix element between spin singlet states. The
resulting expressions for the pionic decay width of the $D_1^*$ state
driven by the axial current operator is then

\begin{equation}
\Gamma_A\left(D_1^*\rightarrow D^*\pi\right)={3\over 8\pi}
{E_{D^*}\over M_{D_1^*}}
\left({g_A^q\over f_\pi}\right)^2 k^3{\cal M}_1^2.
\label{dec2x}
\end{equation}
The corresponding expression for the pion decay width of the spin triplet 
$D_0^*$ state may, using the spin sums given in 
eqs.~(\ref{ss1}) and~(\ref{ss2}), be expressed as

\begin{equation}
\Gamma_A\left(D_0^*\rightarrow D\pi\right)={3\over 8\pi}
{E_D\over M_{D_0^*}}
\left({g_A^q\over f_\pi}\right)^2 k^3{\cal M}_1^2.
\label{dec2xx}
\end{equation}
The axial charge operator also contributes to these decay modes. The
expressions for these $S$-wave contributions may be obtained as

\begin{equation}
\Gamma_C\left(D_1^*\rightarrow D^*\pi\right) = {3\over 64\pi}
{E_{D^*}\over M_{D_1^*}} \left({g_A^q\over f_\pi}\right)^2
\left({\omega_\pi\over m}\right)^2\:k\:{\cal M}_{1S}^2
\label{dec2xxx}
\end{equation}
for the $D_1^*$ meson, and

\begin{equation}
\Gamma_C\left(D_0^*\rightarrow D\pi\right) = {3\over 64\pi}
{E_{D}\over M_{D_0^*}}\left({g_A^q\over f_\pi}\right)^2 
\left({\omega_\pi\over m}\right)^2\:k\:{\cal M}_{1S}^2
\label{dec3xxx}
\end{equation}
for the $D_0^*$ meson. The matrix elements required for the evaluation of
these decay widths are given in Table~\ref{singtab}, and the pion momenta
and masses used can be found in Table~\ref{masstab}. The resulting
calculated pion decay widths of the $D_1^*$ and $D_0^*$ mesons are given
in Table~\ref{singres}. It is evident that the $S$-wave contributions are
dominant in these cases, although the $D$-wave decays also contribute
significantly. In the HQS framework, only $S$-wave pion decay is allowed
to contribute~\cite{Goity}.

\vspace{0.3cm}

\begin{table}[h!]
\begin{center}
\begin{tabular}{l|c|c|c|c|c|c}
\quad Decay & ${\cal M}_1$ & ${\cal M}_1^{\mathrm{ex}}$ &
${\cal M}_{1\mathrm S}$ & ${\cal M}_{1\mathrm S}^{\mathrm{conf}}$ &
${\cal M}_{1\mathrm S}^{\mathrm{OGE}}$ & ${\cal M}_{1\mathrm
S}^{\mathrm{TOT}}$ \\ \hline\hline
$D_1^*\rightarrow D^*\pi$ & 0.086 & 0.010 & -598 MeV & +284 MeV & +135 MeV
& -179 MeV \\ &&&&&& \\
$D_0^*\rightarrow D\pi$   & 0.106 & 0.011 & -578 MeV & +267 MeV & +130 MeV
& -181 MeV 
\end{tabular}
\caption{Matrix elements of the axial current and charge operators for the
decays of the $D_1^*$ and $D_0^*$ mesons. The matrix elements ${\cal
M}_{1\mathrm S}^{\mathrm{TOT}}$ represent the resulting net contribution
to $S$-wave pion decay from the axial charge operator,
eq.~(\ref{onebody}), when the two-body matrix elements from
Section~\ref{conf_sec} are added to the single quark contribution.}
\label{singtab}
\end{center}
\end{table}

\begin{table}[h!]
\begin{center}
\begin{tabular}{l|c|c|c|c}
\quad Decay & Current & Charge & Total & $g_A^q=1$ \\ \hline\hline
$D_1^*\rightarrow D^*\pi$ & 2.8 (2.3) MeV & 7.2 MeV & 10.0 MeV & 13.2 MeV 
\\ &&&& \\
$D_0^*\rightarrow D\pi$ & 7.9 (6.5) MeV & 13.0 MeV & 20.9 MeV & 27.7 MeV 
\end{tabular}
\caption{Predicted pion decay widths of the $D_1^*$ and $D_0^*$ mesons
driven by the axial current and charge operators respectively, for $g_A^q
= 0.87$. The empirical values are total widths~\cite{PDG}, which should
mainly be due to pion decay to the ground state. The numbers in
parentheses are the decay widths obtained without the axial exchange
current contribution. The calculated values are also shown for $g_A^q = 1$.}
\label{singres}
\end{center}
\end{table}

\newpage

The pion widths of the $D_1^*$ and $D_0^*$ states have been predicted to
be much larger in the previous literature~\cite{Goity,Polosa}, although in
ref.~\cite{Goity} a considerable reduction of these widths was already
hinted at. One reason for the smallness of the values obtained here is the
more complete degree of suppression of $S-$wave pion decay modes
achieved here. This suppression was already shown to be necessary in order
to avoid a large overprediction of the width of the $D_1$ meson. Since
eqs.~(\ref{s-decp}) and~(\ref{dec2xxx}) suggest that

\begin{equation}
\Gamma_C\left(D_1^*\rightarrow D^*\pi\right) =  
\frac{3}{2}\:\Gamma_C\left(D_1 \rightarrow D^*\pi\right),
\end{equation}
if the pion momenta $k$ are the same, it immediately follows that the
$S$-wave widths of these two states are of the same order of
magnitude. Consequently, as the triplet $D_1$ state does not have any
large $S$-wave contribution to its pionic width, it follows
that the singlet $D_1^*$ state should be rather narrow as well. 
The calculated pion decay widths of the $D_1^*$ and $D_0^*$ mesons are
here further reduced by the significant spin-orbit splittings in the
$P$-shell, see Fig.~\ref{spektr}. The spectrum obtained in
ref.~\cite{Lahde} predicts the traditional "hydrogen-like" ordering of the
$P$-wave states, which is supported by recent NRQCD lattice
studies~\cite{Wolo}. Since the decay rates are very sensitive to the
amount of phase space available, this effect turns out to be quite
significant. Consequently, the $D_2^*$ meson has considerably more phase
space available than the other $P$-shell states.

In this context, it is instructive to note that the pion decay rates of
the $K_0^*(1430)$ and $K_2^*(1430)$, which are analogous to the $D_0^*$
and $D_2^*$ states, have been experimentally determined~\cite{PDG}. 
The ratio of the widths of the $K\pi$ decays of these strange mesons is
particularly interesting, since the pion momenta in these
decays are exactly equal. The ratio of the width of
$K_0^*\rightarrow K\pi$ to that of $K_2^*\rightarrow K\pi$ is $\sim
5.7$. Using eqs.~(\ref{dec2xx}) and~(\ref{dec3xxx}) 
with the pion momentum and matrix elements of the $D_2^*$ meson, the
corresponding ratio for the $D$ mesons comes to $\sim 5.4$, which supports
the conclusion that the pionic width of the $D_0^*$ state should not be
much larger than $\sim 30$ MeV, as indicated in Table~\ref{singres}.

It should however be noted that as a consequence of the cancellation
between the exchange charge operator contribution and the contribution
from the single quark axial charge operator, the $S$-wave decay widths are
very sensitive to the strength, form and composition of the interaction
Hamiltonian of the $q\bar Q$ system. Empirical determination of the decay
widths of the $D_0^*$ and $D_1^*$ mesons should as a consequence
be able to provide useful information on the forms of these interactions. 

\vspace{0.5cm}

\section{Discussion}

The present calculation of the pion decay widths of the excited
charm mesons indicates that the chiral quark model description is
compatible with the extant experimental data. The underprediction of the
total decay widths of the $L=1$ charm mesons is to be expected,
as the pion decay pattern of the strange meson $L=1$ states as
e.g. the $K_2^*(1430)$, suggests that $\pi\pi$ decay should give a contribution
to the decay width of the $D_2^*$ meson that is about a third as large
as that of single pion decay. 

This point illustrates the difference between the present relativistic
Hamiltonian method and the heavy quark effective field theory
(HQET) method, in which at the present stage the effective coupling strength
of pions to light constituent quarks has to be determined by the 
empirical width of one of the orbitally excited $D$ mesons under the
assumption that the width is solely due to single pion 
decay~\cite{Polosa}. Given that restriction, the present results for the
decay widths of the $D_1$ and $D_2^*$ mesons are overall rather similar to
those in ref.~\cite{Polosa}, although some details as e.g. the ratio
between the decay widths to $D\pi$ and $D^*\pi$ of the $D_2^*$,
eq.~(\ref{ratio}), obtained there is 3.5 and thus somewhat above the
empirical value $2.3\pm 0.6$.

The present calculation,
which was carried out within the framework of the Blankenbecler-Sugar
reduction of the Bethe-Salpeter equation, affirms the observation of
ref.~\cite{Goity}, that a relativistic treatment of the
$q\bar Q$ system is required, as a realistic description
of the pionic decay widths is not attainable with a
non-relativistic description. The framework of the
Blankenbecler-Sugar equation brings the advantage of
formal similarity to the conventional non-relativistic
quantum mechanical treatment. In this approach the
exchange currents that are associated with the scalar confining and vector
one-gluon exchange interactions play an important role in suppressing the
otherwise unrealistically large amplitude for $S-$wave pion decay of the
$D_1$ meson. This also has the effect of drastically reducing the
calculated widths of the singlet $D_1^*$ and triplet $D_0^*$ states, which
are predicted to be extremely broad in the HQS framework. In that
framework, the allowed pion decay modes are determined by the total
angular momentum of the light constituent quark, which is appropriate if
the spins of the light and heavy quarks are decoupled. This is the case if 
the heavy quark mass is very large compared to that of the light
constituent quark. However, the empirical $D$ meson spectrum shown in
Fig.~\ref{spektr} indicates that the spins of the quarks are strongly
coupled, as the $D^*-D$ splitting is of the order $\sim 130$ MeV. In view
of this, the present treatment is more general since it is not
restricted to the case when the heavy quark mass is infinite. The present
work incorporates all decay modes that contribute in that
limit~\cite{Goity}. An analogous treatment within the framework of HQS
requires an $1/M_Q$ expansion along with the introduction of new
parameters, which have to be fitted to experimental data. 

The main goal of studying the pion decays of the charm
mesons is that of determining the numerical value of the axial coupling
$g_A^q$ of the light constituent quarks. As long as the absolute values of
the widths of the $D^*$ mesons are not known, this goal cannot be
conclusively attained. Nevertheless both the present study, as well as the
earlier investigation in ref.~\cite{Goity}, suggests that the standard
values of $g_A^q$, being somewhat below~1 appear to apply well in the
case of the pion decays of the charm mesons. 

\subsubsection*{Acknowledgments}

K.O.E.H. and C.J.N. thank the Finnish Society of Science and Letters for a
stipend. Research supported in part by the Academy of Finland under
contracts 43982 and 44903.

\newpage

\section*{Appendix A}
\label{app}

Because of the smooth radial dependence of the scalar confining and the
one-gluon exchange interactions in the $q\bar Q$ system, the reduced wave
functions of the low-lying states of the $q\bar Q$ system~may be well 
approximated by the following simple expression:

\begin{equation} 
u_l(r)\simeq N_l\:r^l\: e^{-dr^{3/2}}.
\label{appr}
\end{equation} 
Here $m_r$ is the reduced mass of the $q\bar Q$ system and $c$ is the
confining string tension. The coefficient $d$ is defined as

\begin{equation}
d = \frac{2}{3}\sqrt{2m_r c},
\label{fact}
\end{equation}
and $N_l$ is a normalization factor chosen so that

\begin{equation}
\int_0 ^\infty dr\:u_l (r)^2 = 1.
\end{equation}
These approximate expressions have the correct behavior at both small and
large values of the quark separation $r$. For $l=0,1$ the
expressions for the normalization constants may, with the aid of
eq.~(\ref{fact}), be obtained as

\begin{eqnarray}
N_0&=&\frac{4}{3}\sqrt{3m_r c}, \nonumber\\
N_1&=&{2^{11/3}\over 3^{7/3}}{(m_r c)^{5/3}\over \sqrt {\Gamma(10/3)}}.
\end{eqnarray}
The approximate reduced wave functions $u_0(r)$ and $u_1(r)$ are
compared to the corresponding numerical solutions to the 
Blankenbecler-Sugar equation in Fig.~\ref{wavefig}.

\end{document}